\documentclass[nofootinbib,superscriptaddress,notitlepage]{revtex4-1}%
\usepackage{amssymb}
\usepackage{amsfonts}
\usepackage{amsmath}
\usepackage{graphicx}
\usepackage{color}
\usepackage{version}
\usepackage{numinsec}%
\providecommand{\U}[1]{\protect \rule{.1in}{.1in}}

\includeversion{Commentskip}
\includeversion{Commentdel}
\includeversion{Commentnew}
\newcommand{\BSKIP}{\  \color{red}$\Downarrow$SKIP$\Rightarrow$\ }
\newcommand{\ESKIP}{\  \color{red}$\Leftarrow$SKIP$\Uparrow$\color{black}\ }
\newcommand{\BDEL}{\  \color{cyan}$\Downarrow$DEL$\Rightarrow$\ }
\newcommand{\EDEL}{\  \color{cyan}$\Leftarrow$DEL$\Uparrow$\color{black}\ }
\newcommand{\BNEW}{\  \color{blue}$\Downarrow$NEW$\Rightarrow$\ }
\newcommand{\ENEW}{\  \color{blue}$\Leftarrow$NEW$\Uparrow$\color{black}\ }
\begin{document}
\title{Wigner distribution functions for complex dynamical systems: the emergence of
the Wigner-Boltzmann equation}
\author{Dries Sels}
\email{Corresponding author: dries.sels@uatwerpen.be}
\author{Fons Brosens}
\email{fons.brosens@uantwerpen.be}
\affiliation{Physics Department, University of Antwerp, Universiteitsplein 1, 2060
Antwerpen, Belgium}

\begin{abstract}
The equation of motion for the reduced Wigner function of a system coupled to
an external quantum system is presented for the specific case when the
external quantum system can be modeled as a set of harmonic oscillators. The
result is derived from the Wigner function formulation of the Feynman-Vernon
influence functional theory. It is shown how the true self-energy for the
equation of motion is connected with the influence functional for the path
integral. Explicit expressions are derived in terms of the bare Wigner
propagator. Finally, we show under which approximations the resulting equation
of motion reduces to the Wigner-Boltzmann equation.

\end{abstract}
\maketitle

\section{Introduction}

Recently we derived the propagator for the reduced Wigner function of a system
coupled to an external quantum system~\cite{SBMinfluencefunctional}. The
resulting path integral can only be solved exactly for a limited set of
problems. An example of such a system is the Caldeira-Leggett
model~\cite{CaldeiraLeggett} in which the external quantum systems consists of
a set of independent harmonic oscillators bilinearly coupled to the system of
interested. As shown in~\cite{SBMinfluencefunctional}, the resulting time
evolution of the system can be modeled by a stochastic differential equation
subject to correlated noise. In the ohmic case the noise is white and the time
evolution of the system is of the Wigner-Fokker-Planck type, a detailed
discussion of which can be found in~\cite{Kleinert}. In this work we derive
the equation of motion for the reduced Wigner function for a system coupled to
a set of independent oscillators in a slightly more complicated way. We will
still consider the coupling to be linear in the oscillator coordinate but
non-linear in the system coordinate. It is the linearity in the oscillator
coordinate that enables to eliminate them, but the non-linearity in the system
coordinate makes it impossible to solve the resulting path integral by
standard techniques. Note that this situations is quite common as, in the
language of quantum field theory, it describes a particle coupled to a gauge
field. 

The main result of this work can be summarized as follows. Consider a system
and an environment described by the following Hamiltonian%
\begin{equation}
H=\frac{\mathbf{p}^{2}}{2m}+V(\mathbf{x,}t)+\sum_{k}\hbar \omega_{k}\left(
b_{k}^{\dagger}b_{k}+\frac{1}{2}\right)  +\sum_{k}\gamma(\mathbf{k}%
)\exp \left(  i\mathbf{k\cdot x}\right)  b_{k}^{\dagger}+\gamma^{\ast
}(\mathbf{k})\exp \left(  -i\mathbf{k\cdot x}\right)  b_{k}%
.\label{eq:Hamiltonian}%
\end{equation}
Assume furthermore that the phonon bath is initially in thermal equilibrium,
but leave the initial density matrix for the particle unspecified. We shall
show that the Liouville equation for the reduced Wigner function of the system
is then given by%
\begin{equation}
\left(  \frac{\partial}{\partial t}+\frac{\mathbf{p}}{m}\cdot \nabla \right)
f\left(  \mathbf{x},\mathbf{p},t\right)  =\int \mathrm{d}\mathbf{p}^{\prime
}\Lambda(\mathbf{x},\mathbf{p-p}^{\prime},t)f\left(  \mathbf{x},\mathbf{p}%
^{\prime},t\right)  +%
{\displaystyle \iiint}
\mathrm{d}t^{\prime}\mathrm{d}\mathbf{x}^{\prime}\mathrm{d}\mathbf{p}^{\prime
}\Sigma_{R}\left(  \mathbf{x},\mathbf{p},t|\mathbf{x}^{\prime},\mathbf{p}%
^{\prime},t^{\prime}\right)  f\left(  \mathbf{x}^{\prime},\mathbf{p}^{\prime
},t^{\prime}\right)  ,\label{eq:QuantumLiouville}%
\end{equation}
where a retarded self-interaction $\Sigma_{R}$ and a Wigner kernel $\Lambda$
from the external potential are given by
\begin{multline}
\Sigma_{R}\left(  \mathbf{x},\mathbf{p},t|\mathbf{x}^{\prime},\mathbf{p}%
^{\prime},t^{\prime}\right)  =\Theta(t-t^{\prime})\sum_{k}\frac{2\left \vert
\gamma(\mathbf{k})\right \vert ^{2}}{\hbar^{2}}\left[
\begin{array}
[c]{c}%
n_{B}(\omega_{k})\cos \left(  \mathbf{k\cdot}\left(  \mathbf{x}-\mathbf{x}%
^{\prime}\right)  -\omega_{k}\left(  t-t^{\prime}\right)  \right)  \\
+\left(  n_{B}(\omega_{k})+1\right)  \cos \left(  \mathbf{k\cdot}\left(
\mathbf{x}-\mathbf{x}^{\prime}\right)  +\omega_{k}\left(  t-t^{\prime}\right)
\right)
\end{array}
\right]  \times \label{eq:Selfenergy}\\
\times \left[  K_{0}\left(  \mathbf{x},\mathbf{p-}\frac{\hbar \mathbf{k}}%
{2},t|\mathbf{x}^{\prime},\mathbf{p}^{\prime}\mathbf{+}\frac{\hbar \mathbf{k}%
}{2},t^{\prime}\right)  -K_{0}\left(  \mathbf{x},\mathbf{p+}\frac
{\hbar \mathbf{k}}{2},t|\mathbf{x}^{\prime},\mathbf{p}^{\prime}\mathbf{+}%
\frac{\hbar \mathbf{k}}{2},t^{\prime}\right)  \right]
\end{multline}%
\begin{equation}
\Lambda(\mathbf{x},\mathbf{p},t)=-\frac{i}{\hbar}\int \left[  V\left(
\mathbf{x}+\frac{\mathbf{\xi}}{2},t\right)  -V\left(  \mathbf{x}%
-\frac{\mathbf{\xi}}{2},t\right)  \right]  \exp \left(  -\frac{i}{\hbar
}\mathbf{\xi \cdot p}\right)  \frac{\mathrm{d}\mathbf{\xi}}{(2\pi \hbar)^{3}}.
\end{equation}
The complete influence of the bath on the system of interested is contained in
the retarded self-energy $\Sigma_{R}.$ In expression~(\ref{eq:Selfenergy}),
$n_{B}$ denotes the Bose-Einstein distribution at zero chemical potential and
$K_{0}$ denotes the bare Wigner propagator (of the system without the bath).
Most of this manuscript is concerned with the derivation
of~(\ref{eq:QuantumLiouville}) and it is organized in the following way. First
we will derive a perturbation series for reduced Wigner function propagators.
By resumming this series exactly we find a Dyson integral equation for the
reduced propagator. The equation of motion for the reduced Wigner function can
then simply be derived from this integral equation. This will result in an
explicit expression for the retarded self-energy in terms of the bath
properties contained in the influence functional and the bare propagator of
the system. The remaining part of this work will be devoted to a detailed
discussion of the result and its application to translational invariant
systems. It will be shown how the self-energy simplifies to the same
expression as the one predicted by the celebrated Fermi's golden rule under
the usual assumptions of weak-coupling and linear response.

\section{Perturbation theory}

Consider the following propagator%
\[
K_{w}\left(  \mathbf{x}_{b},\mathbf{p}_{b},t_{b}|\mathbf{x}_{a},\mathbf{p}%
_{a},t_{a}\right)  =\frac{1}{\left(  2\pi \hbar \right)  ^{3}}\int
_{\substack{x(t_{a})=x_{a}\\p(t_{a})=m\dot{x}_{a}}}^{\substack{x(t_{b}%
)=x_{b}\\p(t_{b})=m\dot{x}_{b}}}\mathcal{D}\mathbf{x}\int \mathcal{D}%
\mathbf{\xi}\exp \left(  -\frac{i}{\hbar}S_{0}[\mathbf{x,\xi}]+\frac{i}{\hbar
}\Phi \left[  \mathbf{x}+\mathbf{\xi}/2,\mathbf{x}-\mathbf{\xi}/2\right]
\right)
\]
where $S_{0}[\mathbf{x,\xi}]$ is the bare action associated with Hamiltonian
(\ref{eq:Hamiltonian})%
\[
S_{0}[\mathbf{x,\xi}]=\int_{t_{a}}^{t_{b}}m\mathbf{\ddot{x}}(t)\cdot
\mathbf{\xi}(t)+\Delta(\mathbf{x}(t),\mathbf{\xi}(t)\mathbf{,}t)\mathrm{d}t,
\]%
\begin{equation}
\text{with }\Delta(\mathbf{x}(t),\mathbf{\xi}(t)\mathbf{,}t)=V\left(
x\mathbf{+}\frac{\xi}{2},t\right)  -V\left(  x\mathbf{-}\frac{\xi}%
{2},t\right)
\end{equation}
and where $\Phi$ is the influence phase caused by the interactions with the
bath. It was shown in \cite{SBMinfluencefunctional} that, if the bath was
initially in thermal equilibrium, these influence phases are of the form%
\begin{equation}
\frac{i}{\hbar}\Phi \left[  \mathbf{x}+\mathbf{\xi}/2,\mathbf{x}-\mathbf{\xi
}/2\right]  =\int_{t_{a}}^{t_{b}}\int_{t_{a}}^{\tau_{2}}g(\mathbf{x}_{\tau
_{1}},\mathbf{x}_{\tau_{2}},\mathbf{\xi}_{\tau_{1}},\mathbf{\xi}_{\tau_{2}%
})\mathrm{d}\tau_{1}\mathrm{d}\tau_{2},\label{eq:phaseauxfunctiong}%
\end{equation}
where the subindex indicates the time at which the variables must be
evaluated. A short derivation of the influence functional for Hamiltonian
(\ref{eq:Hamiltonian}) will be given in the next section. Clearly the above
path integral is difficult to solve as $g$ is in general a non-linear function
of both $\mathbf{x}$ and $\mathbf{\xi}$ and it contains non-local terms in
time. Hence we proceed by expanding the propagator in a Taylor series around
$g=0$%
\[
K_{w}\left(  \mathbf{x}_{b},\mathbf{p}_{b},t_{b}|\mathbf{x}_{a},\mathbf{p}%
_{a},t_{a}\right)  =\frac{1}{\left(  2\pi \hbar \right)  ^{3}}\int
_{\substack{x(t_{a})=x_{a}\\p(t_{a})=m\dot{x}_{a}}}^{\substack{x(t_{b}%
)=x_{b}\\p(t_{b})=m\dot{x}_{b}}}\mathcal{D}\mathbf{x}\int \mathcal{D}%
\mathbf{\xi}\exp \left(  -\frac{i}{\hbar}S_{0}[\mathbf{x,\xi}]\right)  \sum
_{n}\frac{\left(  \int_{t_{a}}^{t_{b}}\int_{t_{a}}^{\tau_{2}}g(\mathbf{x}%
_{\tau_{1}},\mathbf{x}_{\tau_{2}},\mathbf{\xi}_{\tau_{1}},\mathbf{\xi}%
_{\tau_{2}})\mathrm{d}\tau_{1}\mathrm{d}\tau_{2}\right)  ^{n}}{n!}.
\]
Consider now in more detail the expression for the first order correction
$n=1$%
\[
K_{1}\left(  \mathbf{x}_{b},\mathbf{p}_{b},t_{b}|\mathbf{x}_{a},\mathbf{p}%
_{a},t_{a}\right)  =\int_{t_{a}}^{t_{b}}\mathrm{d}\tau_{2}\int_{t_{a}}%
^{\tau_{2}}\mathrm{d}\tau_{1}\frac{1}{\left(  2\pi \hbar \right)  ^{3}}%
\int_{\substack{x(t_{a})=x_{a}\\p(t_{a})=m\dot{x}_{a}}}^{\substack{x(t_{b}%
)=x_{b}\\p(t_{b})=m\dot{x}_{b}}}\mathcal{D}\mathbf{x}\int \mathcal{D}%
\mathbf{\xi}\exp \left(  -\frac{i}{\hbar}S_{0}[\mathbf{x,\xi}]\right)
g(\mathbf{x}_{\tau_{1}},\mathbf{x}_{\tau_{2}},\mathbf{\xi}_{\tau_{1}%
},\mathbf{\xi}_{\tau_{2}}).
\]
Now we use the Fourier representation of $g,$ i.e.,%
\[
g(\mathbf{x}_{\tau_{1}},\mathbf{x}_{\tau_{2}},\mathbf{\xi}_{\tau_{1}%
},\mathbf{\xi}_{\tau_{2}})=%
{\displaystyle \iint}
g^{\prime}(\mathbf{x}_{\tau_{1}},\mathbf{x}_{\tau_{2}},\mathbf{p}%
,\mathbf{p}^{\prime})\exp \left(  \frac{i}{\hbar}\left(  \mathbf{p}\cdot
\xi_{\tau_{1}}+\mathbf{p}^{\prime}\cdot \xi_{\tau_{2}}\right)  \right)
\mathrm{d}\mathbf{p}\mathrm{d}\mathbf{p}^{\prime},
\]
to arrive at the following expression for $K_{1}$%
\begin{multline*}
K_{1}\left(  \mathbf{x}_{b},\mathbf{p}_{b},t_{b}|\mathbf{x}_{a},\mathbf{p}%
_{a},t_{a}\right)  =\int_{t_{a}}^{t_{b}}\mathrm{d}\tau_{2}\int_{t_{a}}%
^{\tau_{2}}\mathrm{d}\tau_{1}%
{\displaystyle \iint}
\mathrm{d}\mathbf{p}\mathrm{d}\mathbf{p}^{\prime}\frac{1}{\left(  2\pi
\hbar \right)  ^{3}}\int_{\substack{x(t_{a})=x_{a}\\p(t_{a})=m\dot{x}_{a}%
}}^{\substack{x(t_{b})=x_{b}\\p(t_{b})=m\dot{x}_{b}}}\mathcal{D}\mathbf{x}%
\int \mathcal{D}\mathbf{\xi}g^{\prime}(\mathbf{x}_{\tau_{1}},\mathbf{x}%
_{\tau_{2}},\mathbf{p},\mathbf{p}^{\prime})\\
\times \exp \left(  -\frac{i}{\hbar}\left(  S_{0}[\mathbf{x,\xi}]-\int_{t_{a}%
}^{t_{b}}\left(  \mathbf{p}\delta(t-\tau_{1})+\mathbf{p}^{\prime}\delta
(t-\tau_{2})\right)  \cdot \mathbf{\xi}_{t}\mathrm{d}t\right)  \right)  .
\end{multline*}
Clearly the additional terms in the action will cause momentum jumps of size
$\mathbf{p}$ and $\mathbf{p}^{\prime}$ at times $\tau_{1}$ and $\tau_{2}$
respectively. If we furthermore make use of the fact that the zeroth order
propagator satisfies the Chapman-Kolmogorov equation or chain rule then we
directly arrive at%
\begin{multline*}
K_{1}\left(  \mathbf{x}_{b},\mathbf{p}_{b},t_{b}|\mathbf{x}_{a},\mathbf{p}%
_{a},t_{a}\right)  =\int_{t_{a}}^{t_{b}}\mathrm{d}\tau_{2}\int_{t_{a}}%
^{\tau_{2}}\mathrm{d}\tau_{1}%
{\displaystyle \iint}
\mathrm{d}\mathbf{p}\mathrm{d}\mathbf{p}^{\prime}%
{\displaystyle \iint}
\mathrm{d}\mathbf{x}_{2}\mathrm{d}\mathbf{p}_{2}%
{\displaystyle \iint}
\mathrm{d}\mathbf{x}_{1}\mathrm{d}\mathbf{p}_{1}g^{\prime}(\mathbf{x}%
_{1},\mathbf{x}_{2},\mathbf{p},\mathbf{p}^{\prime})\\
\times K_{0}\left(  \mathbf{x}_{b},\mathbf{p}_{b},t_{b}|\mathbf{x}%
_{2},\mathbf{p}_{2},\tau_{2}\right)  K_{0}\left(  \mathbf{x}_{2}%
,\mathbf{p}_{2}-\mathbf{p}^{\prime},\tau_{2}|\mathbf{x}_{1},\mathbf{p}%
_{1}+\mathbf{p},\tau_{1}\right)  K_{0}\left(  \mathbf{x}_{1},\mathbf{p}%
_{1},\tau_{1}|\mathbf{x}_{a},\mathbf{p}_{a},t_{a}\right)  .
\end{multline*}
Now we define%
\begin{equation}
\Sigma(\mathbf{x}_{2},\mathbf{p}_{2},\tau_{2}|\mathbf{x}_{1},\mathbf{p}%
_{1},\tau_{1})=%
{\displaystyle \iint}
\mathrm{d}\mathbf{p}\mathrm{d}\mathbf{p}^{\prime}g^{\prime}(\mathbf{x}%
_{1},\mathbf{x}_{2},\mathbf{p},\mathbf{p}^{\prime})K_{0}\left(  \mathbf{x}%
_{2},\mathbf{p}_{2}-\mathbf{p}^{\prime},\tau_{2}|\mathbf{x}_{1},\mathbf{p}%
_{1}+\mathbf{p},\tau_{1}\right)  ,\label{eq:Selfenergydefiniton}%
\end{equation}
such that%
\[
K_{1}\left(  \mathbf{x}_{b},\mathbf{p}_{b},t_{b}|\mathbf{x}_{a},\mathbf{p}%
_{a},t_{a}\right)  =\int_{t_{a}}^{t_{b}}\mathrm{d}\tau_{2}\int_{t_{a}}%
^{\tau_{2}}\mathrm{d}\tau_{1}%
{\displaystyle \iint}
\mathrm{d}\mathbf{x}_{2}\mathrm{d}\mathbf{p}_{2}%
{\displaystyle \iint}
\mathrm{d}\mathbf{x}_{1}\mathrm{d}\mathbf{p}_{1}K_{0}\left(  \mathbf{B}%
|\mathbf{2}\right)  \Sigma(\mathbf{2}|\mathbf{1})K_{0}\left(  \mathbf{1}%
|\mathbf{A}\right)  ,
\]
where we have introduced the following short hand notation\textbf{
}$\mathbf{J}\mathbb{=}\left \{  \mathbf{x}_{j},\mathbf{p}_{j},\tau_{j}\right \}
.$ Exactly the same procedure can be repeated for the\ subsequent order terms
in the perturbation series. The $n^{th}$ order in the series will have $2n+1$
zeroth order propagators $K_{0}$ and $n$ times $g^{\prime}.$ One thus
constructs $n$ self energies $\Sigma$ which leave $n+1$ zeroth order
propagators $K_{0}$ to connect all self energies. Time ordering all the
unordered time integrals will exactly cancel the $n!$ term in the denominator.
Such a structure immediately gives rise to a recurrence relation, from which
we find the following Dyson integral equation for the propagator
\begin{equation}
K_{w}\left(  \mathbf{B}|\mathbf{A}\right)  =K_{0}\left(  \mathbf{B}%
|\mathbf{A}\right)  +\int_{t_{a}}^{t_{b}}\mathrm{d}\tau_{2}\int_{t_{a}}%
^{\tau_{2}}\mathrm{d}\tau_{1}%
{\displaystyle \iint}
\mathrm{d}\mathbf{x}_{2}\mathrm{d}\mathbf{p}_{2}%
{\displaystyle \iint}
\mathrm{d}\mathbf{x}_{1}\mathrm{d}\mathbf{p}_{1}K_{0}\left(  \mathbf{B}%
|\mathbf{2}\right)  \Sigma(\mathbf{2}|\mathbf{1})K_{w}\left(  \mathbf{1}%
|\mathbf{A}\right)  .\label{eq:Dyson}%
\end{equation}
This constitutes the main result of the section. Before we finally turn our
attention to the expression for the self-energy, we derive
Eq.~(\ref{eq:QuantumLiouville}) from the Dyson equation above. Since $K_{0}$
is just the bare propagator associated with the bare action $S_{0}%
[\mathbf{x},\mathbf{\xi}],$ it satisfies the Wigner-Liouville
equation~\cite{SBMpathintegral}%
\[
\left(  \frac{\partial}{\partial t_{b}}+\frac{\mathbf{p}_{b}}{m}\cdot
\nabla \right)  K_{0}\left(  \mathbf{x}_{b},\mathbf{p}_{b},t_{b}|\mathbf{x}%
_{a},\mathbf{p}_{a},t_{a}\right)  =\int \mathrm{d}\mathbf{p}^{\prime}%
\Lambda(\mathbf{x}_{b},\mathbf{p}_{b}\mathbf{-p}^{\prime},t_{b})K_{0}\left(
\mathbf{x}_{b},\mathbf{p}^{\prime},t_{b}|\mathbf{x}_{a},\mathbf{p}_{a}%
,t_{a}\right)  ,
\]
which we denote as $\partial_{t_{b}}K_{0}\left(  \mathbf{B}|\mathbf{A}\right)
+\hat{L}_{b}K_{0}\left(  \mathbf{B}|\mathbf{A}\right)  =0.$ Differentiating
Eq.~(\ref{eq:Dyson}) with respect to the final time $t_{b}$ yields%
\begin{multline}
\frac{\partial}{\partial t_{b}}K_{w}\left(  \mathbf{B}|\mathbf{A}\right)
=\frac{\partial}{\partial t_{b}}K_{0}\left(  \mathbf{B}|\mathbf{A}\right)
+\int_{t_{a}}^{t_{b}}\mathrm{d}\tau_{2}\int_{t_{a}}^{\tau_{2}}\mathrm{d}%
\tau_{1}%
{\displaystyle \iint}
\mathrm{d}\mathbf{x}_{2}\mathrm{d}\mathbf{p}_{2}%
{\displaystyle \iint}
\mathrm{d}\mathbf{x}_{1}\mathrm{d}\mathbf{p}_{1}\frac{\partial}{\partial
t_{b}}K_{0}\left(  \mathbf{B}|\mathbf{2}\right)  \Sigma(\mathbf{2}%
|\mathbf{1})K_{w}\left(  \mathbf{1}|\mathbf{A}\right)  \nonumber \\
+\int_{t_{a}}^{t_{b}}\mathrm{d}\tau_{1}%
{\displaystyle \iint}
\mathrm{d}\mathbf{x}_{2}\mathrm{d}\mathbf{p}_{2}%
{\displaystyle \iint}
\mathrm{d}\mathbf{x}_{1}\mathrm{d}\mathbf{p}_{1}K_{0}\left(  \mathbf{B}%
|\mathbf{x}_{2},\mathbf{p}_{2},t_{b}\right)  \Sigma(\mathbf{x}_{2}%
,\mathbf{p}_{2},t_{b}|\mathbf{1})K_{w}\left(  \mathbf{1}|\mathbf{A}\right)  .
\end{multline}
In the last line we now find a propagator where the initial and final times
are equal. By definition of the propagator this yields%
\[
K_{0}\left(  \mathbf{B}|\mathbf{x}_{2},\mathbf{p}_{2},t_{b}\right)
=\delta(\mathbf{x}_{b}-\mathbf{x}_{2})\delta \left(  \mathbf{p}_{b}%
-\mathbf{p}_{2}\right)  .
\]
If we furthermore introduce $\hat{L}_{b}$ and take it out of the integral we
find%
\begin{multline}
\frac{\partial}{\partial t_{b}}K_{w}\left(  \mathbf{B}|\mathbf{A}\right)
=-\hat{L}_{b}\left[  K_{0}\left(  \mathbf{B}|\mathbf{A}\right)  +\int_{t_{a}%
}^{t_{b}}\mathrm{d}\tau_{2}\int_{t_{a}}^{\tau_{2}}\mathrm{d}\tau_{1}%
{\displaystyle \iint}
\mathrm{d}\mathbf{x}_{2}\mathrm{d}\mathbf{p}_{2}%
{\displaystyle \iint}
\mathrm{d}\mathbf{x}_{1}\mathrm{d}\mathbf{p}_{1}K_{0}\left(  \mathbf{B}%
\mathbb{|}\mathbf{2}\right)  \Sigma(\mathbf{2}|\mathbf{1})K_{w}\left(
\mathbf{1}|\mathbf{A}\right)  \right]  \nonumber \\
+\int_{t_{a}}^{t_{b}}\mathrm{d}\tau_{1}%
{\displaystyle \iint}
\mathrm{d}\mathbf{x}_{1}\mathrm{d}\mathbf{p}_{1}\Sigma(\mathbf{B}%
|\mathbf{1})K_{w}\left(  \mathbf{1}|\mathbf{A}\right)  .
\end{multline}
The term between the square brackets is exactly the right hand side of
Eq.~(\ref{eq:Dyson}), such that we arrive at%
\begin{equation}
\frac{\partial}{\partial t_{b}}K_{w}\left(  \mathbf{B}|\mathbf{A}\right)
+\hat{L}_{b}K_{w}\left(  \mathbf{B}|\mathbf{A}\right)  =\int_{t_{a}}^{t_{b}%
}\mathrm{d}\tau_{1}%
{\displaystyle \iint}
\mathrm{d}\mathbf{x}_{1}\mathrm{d}\mathbf{p}_{1}\Sigma(\mathbf{B}%
|\mathbf{1})K_{w}\left(  \mathbf{1}|\mathbf{A}\right)  .
\end{equation}
If we multiply the expression with the initial reduced Wigner function and
integrate over $\left(  \mathbf{x}_{a},\mathbf{p}_{a}\right)  $ we find the
equation of motion for the reduced Wigner function%
\[
\frac{\partial}{\partial t_{b}}f\left(  \mathbf{B}\right)  +\hat{L}%
_{b}f\left(  \mathbf{B}\right)  =\int_{t_{a}}^{t_{b}}\mathrm{d}\tau_{1}%
{\displaystyle \iint}
\mathrm{d}\mathbf{x}_{1}\mathrm{d}\mathbf{p}_{1}\Sigma(\mathbf{B}%
|\mathbf{1})f(\mathbf{1}).
\]
At this point it is safe to take the limit of $t_{a}\rightarrow-\infty,$ such
that the assumption of an initial product state of bath and particle was
infinitely long ago. Defining the retarded self-energy as%
\[
\Sigma_{R}\left(  \mathbf{2|1}\right)  =\Theta(t_{2}-t_{1})\Sigma \left(
\mathbf{2|1}\right)  ,
\]
and replacing $\hat{L}_{b}$ with its specific expression we finally arrive at
the quantum Liouville equation~(\ref{eq:QuantumLiouville}) announced in the
introduction:%
\begin{equation}
\left(  \frac{\partial}{\partial t}+\frac{\mathbf{p}}{m}\cdot \nabla \right)
f\left(  \mathbf{x},\mathbf{p},t\right)  =\int \mathrm{d}\mathbf{p}^{\prime
}\Lambda(\mathbf{x},\mathbf{p-p}^{\prime},t)f\left(  \mathbf{x},\mathbf{p}%
^{\prime},t\right)  +%
{\displaystyle \iiint}
\mathrm{d}t^{\prime}\mathrm{d}\mathbf{x}^{\prime}\mathrm{d}\mathbf{p}^{\prime
}\Sigma_{R}\left(  \mathbf{x},\mathbf{p},t|\mathbf{x}^{\prime},\mathbf{p}%
^{\prime},t^{\prime}\right)  f\left(  \mathbf{x}^{\prime},\mathbf{p}^{\prime
},t^{\prime}\right)  .
\end{equation}

\section{Influence functional and self-energy}

In this section we show how to arrive at expression~(\ref{eq:Selfenergy}) or
the retarded self-energy $\Sigma_{R}$ from its basic
definition~(\ref{eq:Selfenergydefiniton}). To do so we need the influence
functional~\cite{FeyVer, SBMinfluencefunctional} caused by the bath. A
detailed derivation of the influence functional of a harmonic subsystem
linearly coupled to the system was presented in
reference~\cite{SBMinfluencefunctional}. Representing the bosonic bath
operator in terms of harmonic oscillator ladder operator directly turns the
action associated with Hamiltonian~(\ref{eq:Hamiltonian}) into the form
discussed in~\cite{SBMinfluencefunctional}. Here we present a much shorter
derivation, using coherent state Wigner functions~\cite{Polkovnikov} for the
bath. The Weyl ordered Lagrangian associated with
Hamiltonian~(\ref{eq:Hamiltonian}) is%
\begin{multline*}
\mathcal{L}_{W}\left(  \mathbf{x,\dot{x},}\left \{  b_{k}\right \}  ,\left \{
b_{k}^{\ast}\right \}  \right)  =\frac{m\mathbf{\dot{x}}^{2}}{2}+\frac{1}%
{2}\sum_{k}\left(  b_{k}^{\ast}i\hbar \frac{\partial b_{k}}{\partial t}%
-b_{k}i\hbar \frac{\partial b_{k}^{\ast}}{\partial t}\right)  \\
-V(\mathbf{x,}t)-\sum_{k}\hbar \omega_{k}b_{k}^{\ast}b_{k}-\sum_{k}%
\gamma(\mathbf{k})\exp \left(  i\mathbf{k\cdot x}\right)  b_{k}^{\ast}%
+\gamma^{\ast}(\mathbf{k})\exp \left(  -i\mathbf{k\cdot x}\right)  b_{k.}%
\end{multline*}
Since the action is harmonic in the bath variables its Wigner propagator is
just a delta function along the classical trajectory~\cite{Polkovnikov,
SBMinfluencefunctional,Elze}. Hence in complete analogy with the derivation in
\cite{SBMinfluencefunctional}, we find the following expression for the
influence functional%
\begin{multline*}
\mathcal{F}\left[  \mathbf{x+\xi}/2,\mathbf{x-\xi}/2\right]  =\exp \left(
\frac{i}{\hbar}\sum_{k}\frac{4\left \vert \gamma(\mathbf{k})\right \vert ^{2}%
}{\hbar}\int_{t_{a}}^{t_{b}}\int_{t_{a}}^{t}\sin(\mathbf{k\cdot}\left(
\mathbf{x}_{s}-\mathbf{x}_{t}\right)  )\sin \left(  \frac{\mathbf{k\cdot \xi
}_{t}}{2}\right)  \cos \left(  \frac{\mathbf{k\cdot \xi}_{s}}{2}\right)
\sin(\omega_{k}\left[  t-s\right]  )\text{d}s\text{d}t\right)  \\
\times \prod_{k}\left \{
{\displaystyle \iint}
\mathrm{d}b_{k}\mathrm{d}b_{k}^{\ast}\exp \left(  \frac{2\gamma(\mathbf{k}%
)}{\hbar}b_{k}^{\ast}\int_{t_{a}}^{t_{b}}e^{i\omega_{k}(t-t_{a})+i\mathbf{kx}%
_{t}}\sin \left(  \frac{\mathbf{k\cdot \xi}_{t}}{2}\right)  \mathrm{d}%
t-h.c.\right)  f_{k}(b_{k},b_{k}^{\ast})\right \}  .
\end{multline*}
In the case of an initial thermal distribution with inverse temperature
$\beta$ we have%
\[
f_{k}(b_{k},b_{k}^{\ast})=\frac{2\tanh \left(  \beta \hbar \omega_{k}/2\right)
}{\pi}\exp \left(  -2\tanh \left(  \beta \hbar \omega_{k}/2\right)  \left \vert
b_{k}\right \vert ^{2}\right)  .
\]
The remaining Gaussian integral then results in the following expression for
the influence phase%
\begin{multline*}
\Phi \left[  \mathbf{x}+\mathbf{\xi}/2,\mathbf{x}-\mathbf{\xi}/2\right]
=\sum_{k}\frac{4\left \vert \gamma(\mathbf{k})\right \vert ^{2}}{\hbar}%
\int_{t_{a}}^{t_{b}}\int_{t_{a}}^{t}\sin(\mathbf{k\cdot}\left(  \mathbf{x}%
_{s}-\mathbf{x}_{t}\right)  )\sin \left(  \frac{\mathbf{k\cdot \xi}_{t}}%
{2}\right)  \cos \left(  \frac{\mathbf{k\cdot \xi}_{s}}{2}\right)  \sin
(\omega_{k}\left[  t-s\right]  )\text{d}s\text{d}t\\
+i\sum_{k}\frac{\coth \left(  \frac{\beta \hbar \omega_{k}}{2}\right)  }{2}%
\frac{4\left \vert \gamma(\mathbf{k})\right \vert ^{2}}{\hbar}\int_{t_{a}%
}^{t_{b}}\int_{t_{a}}^{t_{b}}\cos(\mathbf{k\cdot}\left(  \mathbf{x}%
_{s}-\mathbf{x}_{t}\right)  )\sin \left(  \frac{\mathbf{k\cdot \xi}_{t}}%
{2}\right)  \sin \left(  \frac{\mathbf{k\cdot \xi}_{s}}{2}\right)  \cos
(\omega_{k}\left[  t-s\right]  )\text{d}s\text{d}t.
\end{multline*}
Time ordering of the last term removes the factor $1/2$ and brings the
influence phase in the desired form~(\ref{eq:phaseauxfunctiong}), which allows
to extract $g(\mathbf{x}_{t},\mathbf{x}_{s},\mathbf{\xi}_{t},\mathbf{\xi}%
_{s})$ from the previous expression for $\Phi.$ From
definition~(\ref{eq:Selfenergydefiniton}) one arrives by straightforward
algebra at the final expression~(\ref{eq:Selfenergy}) for the self-energy
$\Sigma_{R}$, which we recapitulate here for convenience%
\begin{multline*}
\Sigma_{R}\left(  \mathbf{x},\mathbf{p},t|\mathbf{x}^{\prime},\mathbf{p}%
^{\prime},t^{\prime}\right)  =\Theta(t-t^{\prime})\sum_{k}\frac{2\left \vert
\gamma(\mathbf{k})\right \vert ^{2}}{\hbar^{2}}\left[
\begin{array}
[c]{c}%
n_{B}(\omega_{k})\cos \left(  \mathbf{k\cdot}\left(  \mathbf{x}-\mathbf{x}%
^{\prime}\right)  -\omega_{k}\left(  t-t^{\prime}\right)  \right)  \\
+\left(  n_{B}(\omega_{k})+1\right)  \cos \left(  \mathbf{k\cdot}\left(
\mathbf{x}-\mathbf{x}^{\prime}\right)  +\omega_{k}\left(  t-t^{\prime}\right)
\right)
\end{array}
\right]  \times \\
\times \left[  K_{0}\left(  \mathbf{x},\mathbf{p-}\frac{\hbar \mathbf{k}}%
{2},t|\mathbf{x}^{\prime},\mathbf{p}^{\prime}\mathbf{+}\frac{\hbar \mathbf{k}%
}{2},t^{\prime}\right)  -K_{0}\left(  \mathbf{x},\mathbf{p+}\frac
{\hbar \mathbf{k}}{2},t|\mathbf{x}^{\prime},\mathbf{p}^{\prime}\mathbf{+}%
\frac{\hbar \mathbf{k}}{2},t^{\prime}\right)  \right]  .
\end{multline*}
One readily identifies the scattering terms associated with emission and
absorption of quanta from the reservoir. Moreover note that, independent of
the bare Hamiltonian of the system,
\[
\int \mathrm{d}\mathbf{p}\Sigma_{R}\left(  \mathbf{x},\mathbf{p},t|\mathbf{x}%
^{\prime},\mathbf{p}^{\prime},t^{\prime}\right)  =0,
\]
which implies that the reduced distribution function satisfies the continuity equation.

\section{Discussion}

As an important example, consider the bare system to be free, i.e.
$H_{0}=\mathbf{p}^{2}/2m.$ Since the problem is harmonic its Wigner propagator
is determined by the classical trajectory%
\[
K_{0}\left(  \mathbf{x},\mathbf{p},t|\mathbf{x}^{\prime},\mathbf{p}^{\prime
},t^{\prime}\right)  =\delta(\mathbf{p-p}^{\prime})\delta \left(
\mathbf{x}-\left(  \mathbf{x}^{\prime}+\frac{\mathbf{p}^{\prime}}%
{m}(t-t^{\prime})\right)  \right)  .
\]
Note that for a bare free particle both the bare and the total Hamiltonian of
system and bath are translational invariant. Let us therefore consider the
initial reduced Wigner function of the system to be translational invariant
too, i.e., $f_{0}(\mathbf{x,p})=f(\mathbf{p})$. Under these conditions the
Liouville equation (\ref{eq:QuantumLiouville}) for the reduced Wigner function
becomes
\begin{multline}
\frac{\partial}{\partial t}f\left(  \mathbf{p},t\right)  =\sum_{k}%
\frac{2\left \vert \gamma(\mathbf{k})\right \vert ^{2}}{\hbar^{2}}\int
\mathrm{d}s\Theta(t-s)\left[
\begin{array}
[c]{c}%
n_{B}(\omega_{k})\cos \left(  \left[  \frac{\left(  \mathbf{p}+\hbar
\mathbf{k}\right)  ^{2}}{2m}-\frac{\mathbf{p}^{2}}{2m}+\hbar \omega_{k}\right]
\left(  t-s\right)  /\hbar \right)  \\
+\left(  n_{B}(\omega_{k})+1\right)  \cos \left(  \left[  \frac{\left(
\mathbf{p}+\hbar \mathbf{k}\right)  ^{2}}{2m}-\frac{\mathbf{p}^{2}}{2m}%
-\hbar \omega_{k}\right]  \left(  t-s\right)  /\hbar \right)
\end{array}
\right]  f\left(  \mathbf{p+\hbar \mathbf{k}},s\right)  \\
-\sum_{k}\frac{2\left \vert \gamma(\mathbf{k})\right \vert ^{2}}{\hbar^{2}}%
\int \mathrm{d}s\Theta(t-s)\left[
\begin{array}
[c]{c}%
n_{B}(\omega_{k})\cos \left(  \left[  \frac{\left(  \mathbf{p}+\hbar
\mathbf{k}\right)  ^{2}}{2m}-\frac{\mathbf{p}^{2}}{2m}-\omega_{k}\right]
\left(  t-s\right)  /\hbar \right)  \\
+\left(  n_{B}(\omega_{k})+1\right)  \cos \left(  \left[  \frac{\left(
\mathbf{p}+\hbar \mathbf{k}\right)  ^{2}}{2m}-\frac{\mathbf{p}^{2}}{2m}%
+\omega_{k}\right]  \left(  t-s\right)  /\hbar \right)
\end{array}
\right]  f\left(  \mathbf{p},s\right)  .
\end{multline}
Consequently, any stationary solution of the Liouville equation should satisfy%
\begin{equation}
\sum_{k}\frac{2\pi \left \vert \gamma(\mathbf{k})\right \vert ^{2}}{\hbar}\left[
\begin{array}
[c]{c}%
n_{B}(\omega_{k})S_{+}\left(  \mathbf{p,k,}\omega_{k}\right)  \\
+\left(  n_{B}(\omega_{k})+1\right)  S_{-}\left(  \mathbf{p,k,}\omega
_{k}\right)
\end{array}
\right]  f\left(  \mathbf{p+\hbar \mathbf{k}}\right)  =\sum_{k}\frac
{2\pi \left \vert \gamma(\mathbf{k})\right \vert ^{2}}{\hbar}\left[
\begin{array}
[c]{c}%
n_{B}(\omega_{k})S_{-}\left(  \mathbf{p,k,}\omega_{k}\right)  \\
+\left(  n_{B}(\omega_{k})+1\right)  S_{+}\left(  \mathbf{p,k,}\omega
_{k}\right)
\end{array}
\right]  f\left(  \mathbf{p}\right)  ,\label{eq:Freeparticlegoldenrule}%
\end{equation}
with%
\[
S_{\pm}\left(  \mathbf{p,k,}\omega_{k}\right)  =\int \frac{\mathrm{d}s}%
{\pi \hbar}\Theta(t-s)\cos \left(  \left[  \frac{\left(  \mathbf{p}%
+\hbar \mathbf{k}\right)  ^{2}}{2m}-\frac{\mathbf{p}^{2}}{2m}\pm \hbar \omega
_{k}\right]  \left(  t-s\right)  /\hbar \right)  =\delta \left(  \frac{\left(
\mathbf{p}+\hbar \mathbf{k}\right)  ^{2}}{2m}-\frac{\mathbf{p}^{2}}{2m}\pm
\hbar \omega_{k}\right)  .
\]
Clearly the transition rates are now determined by Fermi's golden rule.
Moreover Eq.~(\ref{eq:Freeparticlegoldenrule}) now manifests detailed balance,
from which we immediately find that
\[
\frac{f\left(  \mathbf{p+\hbar \mathbf{k}}\right)  }{f\left(  \mathbf{p}%
\right)  }=\frac{n_{B}\left(  \frac{\left(  \mathbf{p}+\hbar \mathbf{k}\right)
^{2}}{2m}-\frac{\mathbf{p}^{2}}{2m}\right)  }{n_{B}\left(  \frac{\left(
\mathbf{p}+\hbar \mathbf{k}\right)  ^{2}}{2m}-\frac{\mathbf{p}^{2}}{2m}\right)
+1}=\exp \left(  -\beta \left(  \frac{\left(  \mathbf{p}+\hbar \mathbf{k}\right)
^{2}}{2m}-\frac{\mathbf{p}^{2}}{2m}\right)  \right)  .
\]
This implies that the only stationary reduced Wigner function for the system
is given by a Maxwell-Boltzmann distribution. Note that this result is
independent of the strength of the coupling $\gamma$ and the nature of the
bath $\omega_{k}.$ With this in mind, we perturb the system by adding an
external force such that
\[
H_{0}=\frac{\mathbf{p}^{2}}{2m}-\mathbf{F\cdot x,}%
\]
Since the bare system Hamiltonian remains quadratic we find
\[
K_{0}\left(  \mathbf{x},\mathbf{p},t|\mathbf{x}^{\prime},\mathbf{p}^{\prime
},t^{\prime}\right)  =\delta \left(  \mathbf{p-}\left(  \mathbf{p}^{\prime
}+\mathbf{F}\left(  t-t^{\prime}\right)  \right)  \right)  \delta \left(
\mathbf{x}-\left(  \mathbf{x}^{\prime}+\frac{\mathbf{p}^{\prime}}%
{m}(t-t^{\prime})+\frac{\mathbf{F}}{2m}\left(  t-t^{\prime}\right)
^{2}\right)  \right)  .
\]
If we again consider the initial distribution to be homogeneous, we arrive at
the following Liouville equation for the perturbed system%
\begin{multline}
\left(  \frac{\partial}{\partial t}+\mathbf{F\cdot}\nabla_{p}\right)  f\left(
\mathbf{p},t\right)  =\sum_{k}\frac{2\left \vert \gamma(\mathbf{k})\right \vert
^{2}}{\hbar^{2}}\int \mathrm{d}s\Theta(s)\left[
\begin{array}
[c]{c}%
n_{B}(\omega_{k})\cos \left(  \Delta E_{+}s/\hbar-\frac{\mathbf{k\cdot F}}%
{2m}s^{2}\right)  \\
+\left(  n_{B}(\omega_{k})+1\right)  \cos \left(  \Delta E_{-}s/\hbar
-\frac{\mathbf{k\cdot F}}{2m}s^{2}\right)
\end{array}
\right]  f\left(  \mathbf{p+\hbar k-F}s,t-s\right)  \\
-\sum_{k}\frac{2\left \vert \gamma(\mathbf{k})\right \vert ^{2}}{\hbar^{2}}%
\int \mathrm{d}s\Theta(s)\left[
\begin{array}
[c]{c}%
n_{B}(\omega_{k})\cos \left(  \Delta E_{-}s/\hbar-\frac{\mathbf{k\cdot F}}%
{2m}s^{2}\right)  \\
+\left(  n_{B}(\omega_{k})+1\right)  \cos \left(  \Delta E_{+}s/\hbar
-\frac{\mathbf{k\cdot F}}{2m}s^{2}\right)
\end{array}
\right]  f\left(  \mathbf{p-F}s,t-s\right)  ,
\end{multline}
where we introduced $\Delta E_{\pm}\left(  \mathbf{p,k,}\omega_{k}\right)
=\frac{\left(  \mathbf{p}+\hbar \mathbf{k}\right)  ^{2}}{2m}-\frac
{\mathbf{p}^{2}}{2m}\pm \hbar \omega_{k}$ for notational simplicity. The right
hand side of this equation is highly non-Markovian and difficult to treat.
However, note that as a result of the bare propagator $K_{0}$ in the
expression of the self-energy the Wigner function should be evaluated at a
time $s$ before the current time while also being displaced along the
classical trajectory. In the absence of the bath, this would exactly yield the
Wigner function at the current time. Hence under weak coupling conditions we
have%
\begin{align*}
f\left(  \mathbf{p-F}s,t-s\right)   &  \approx f\left(  \mathbf{p},t\right)
+\mathcal{O}\left(  \left \vert \gamma(\mathbf{k})\right \vert ^{2}\right)  ,\\
f\left(  \mathbf{p+\hbar k-F}s,t-s\right)   &  \approx f\left(
\mathbf{p+\hbar k},t\right)  +\mathcal{O}\left(  \left \vert \gamma
(\mathbf{k})\right \vert ^{2}\right)  .
\end{align*}
As the right hand side of the equation of motion already contains a factor
$\left \vert \gamma(\mathbf{k})\right \vert ^{2}$ we can neglect the correction
to the classical trajectory in weak coupling. This approximation results in
the following Markovian Liouville equation%
\begin{multline}
\left(  \frac{\partial}{\partial t}+\mathbf{F\cdot}\nabla_{p}\right)  f\left(
\mathbf{p},t\right)  \approx \sum_{k}\frac{2\pi \left \vert \gamma(\mathbf{k}%
)\right \vert ^{2}}{\hbar}\left[
\begin{array}
[c]{c}%
n_{B}(\omega_{k})S_{+}\left(  \mathbf{p,k,}\omega_{k}\right)  \\
+\left(  n_{B}(\omega_{k})+1\right)  S_{-}\left(  \mathbf{p,k,}\omega
_{k}\right)
\end{array}
\right]  f\left(  \mathbf{p+\hbar k},t\right)  \\
-\sum_{k}\frac{2\pi \left \vert \gamma(\mathbf{k})\right \vert ^{2}}{\hbar
}\left[
\begin{array}
[c]{c}%
n_{B}(\omega_{k})S_{-}\left(  \mathbf{p,k,}\omega_{k}\right)  \\
+\left(  n_{B}(\omega_{k})+1\right)  S_{+}\left(  \mathbf{p,k,}\omega
_{k}\right)
\end{array}
\right]  f\left(  \mathbf{p},t\right)  ,
\end{multline}
where
\[
S_{\pm}\left(  \mathbf{p,k,}\omega_{k}\right)  =\int_{0}^{\infty}%
\frac{\mathrm{d}\beta}{\pi}\cos \left(  \beta \Delta E_{\pm}-\frac{\hbar
^{2}\mathbf{k\cdot F}}{2m}\beta^{2}\right)  .
\]
It follows immediately that under linear response conditions, i.e., when the
perturbation is weak, the scattering again becomes%
\[
\lim_{\mathbf{F}\rightarrow0}S_{\pm}\left(  \mathbf{p,k,}\omega_{k}\right)
=\delta \left(  \frac{\left(  \mathbf{p}+\hbar \mathbf{k}\right)  ^{2}}%
{2m}-\frac{\mathbf{p}^{2}}{2m}\pm \hbar \omega_{k}\right)  ,
\]
and the reduced Liouville equation the system simply becomes a classical
Boltzmann equation where the scattering rates are given by Fermi's golden
rule. Moreover note that the force dependent term in $S_{\pm}$ is quantum in
nature as it scales with $\hbar^{2}.$ In this respect the reduction of the
scattering amplitudes to Fermi's golden rule under the assumption of
vanishingly small perturbation can also be interpreted as a truncation of the
exact quantum result up to the classical result. Similarly one can, for an
isolated system, truncate the Moyal bracket up to the Poisson bracket, which
also results in quantum correction of order $\hbar^{2}$ because the Poisson
bracket does not incorporate any interference effects in the time evolution.
In this context this approximation is known as the truncated Wigner
approximation~\cite{Polkovnikov}. The $\beta$ integral which determines the
scattering amplitude can, in this case, be done exactly as it is just a
complex Gaussian integral. Doing so one finds that the resulting function
tends to a delta function in a rather complicated way. On the one hand,
depending on the sign of $\mathbf{k\cdot F,}$ the scattering amplitude
oscillates rapidly, while it decays monotonically on the other hand. This
oscillating behavior of the scattering amplitude again indicates the quantum
nature of the scattering as it makes the scattering amplitude negative for
some (non classical) transitions. This in turn allows the Wigner function to
become negative. Finally note that, although a single scattering event does
not have to preserve energy anymore, the expected energy difference is still
zero as the first moment of $S_{\pm}$ with respect to $\Delta E_{\pm}$ vanishes.

\section{Conclusion}

This work introduces the Wigner-Liouville equation for the reduced
distribution function of a system coupled to a bath of harmonic oscillators.
It shows how the retarded self-energy can be calculated from the influence
functional of the reduced propagator. In general, the resulting Liouville
equation is highly non-Markovian. How to deal with it is still the subject of
current research. However, it was shown that under the conditions of weak
coupling the time evolution approximately becomes Markovian. An additional
assumption of linear response moreover results in Fermi's golden rule for the
scattering rates. Alternatively, one can interpret Fermi's golden rule as the
classically truncated version of the exact quantum amplitudes. Whereas the
exact amplitudes include events which change the sign of the Wigner
distribution, Fermi's golden rule does not. Finally it was shown that,
whenever the bare system is a free particle, the stationary distribution
satisfies detailed balance conditions and is consequently given by a
Maxwell-Boltzmann distribution.

%

%

\end{document}